\documentclass[review]{elsarticle}
\usepackage{hyperref}
\usepackage[flushleft]{threeparttable}
\usepackage{lscape}
\usepackage[version=4]{mhchem}
\usepackage{isotope}

\journal{Journal of \LaTeX\ Templates}

\begin{document}

\begin{frontmatter}

\title{Nuclear Isotope Production by Ordinary Muon Capture Reaction \tnoteref{mytitlenote}}
\tnotetext[mytitlenote]{This project is partially funded through the UTM-International Industry Institute Grant (IIIG) Grant No Q.J130000.3026.01M14 and the Fundamental Research Grant Scheme (FRGS) Grant No. R.J130000.7826.4F971 of the Ministry of Higher Education and University Teknologi Malaysia. The authors thank Prof Dr Yoshitaka Kuno and MuSIC colleagues at Osaka University and Billion Prima Sdn. Bhd. for their valuable contribution.}


\author[mymainaddress,mysecondaryaddress,mythirdaddress]{I.H. Hashim\corref{mycorrespondingauthor}}
\cortext[mycorrespondingauthor]{Corresponding author}
\ead[url]{izyan@utm.my}

\author[mysecondaryaddress,myforthaddress]{H. Ejiri}
\author[mymainaddress]{F. Othman}
\author[mymainaddress]{F. Ibrahim}
\author[mymainaddress]{F. Soberi}
\author[mymainaddress]{N.N.A.M.A. Ghani}
\author[mysecondaryaddress]{T. Shima}
\author[myfifthaddress]{A. Sato}
\author[mysixthaddress]{K. Ninomiya}

\address[mymainaddress]{Department of Physics, Faculty of Science, Universiti Teknologi Malaysia, 81310 Johor Bahru, Johor, Malaysia.}
\address[mysecondaryaddress]{Research Center for Nuclear Physics, Osaka University, Ibaraki Osaka 567-0047, Japan.}
\address[mythirdaddress]{National Centre for Particle Physics, Universiti Malaya, 50603 Kuala Lumpur, Malaysia}
\address[myforthaddress]{Nuclear Science, Czech Technical University, Prague, Czech Republic.}
\address[myfifthaddress]{Department of Physics, Osaka University, Toyonaka Osaka 560-0043, Japan.}
\address[mysixthaddress]{Department of Chemistry, Osaka University, Toyonaka Osaka 560-0043, Japan.}

\begin{abstract}
Muon capture isotope production (MuCIP) using negative ordinary muon ($\mu$) capture reactions (OMC) is used to efficiently produce various kinds of nuclear isotopes for both fundamental and applied science studies.
The large capture probability of $\mu$ into a nucleus, together with the high intensity $\mu$ beam, make it possible to produce nuclear isotopes in the order of 10 $^{9-10}$ per second depending on the muon beam intensity. 
Radioactive isotopes (RIs) produced by MuCIP are complementary to those produced by photon and neutron capture reactions and are used for various science and technology applications. 
MuCIP on \ce{^{Nat}}Mo by using the RCNP MuSIC $\mu$ beam is presented to demonstrate the feasibility of MuCIP.
Nuclear isotopes produced by MuCIP are evaluated by using a pre-equilibrium (PEQ) and equilibrium (EQ) proton neutron emission model. 
Radioactive $^{99}$Mo isotopes and the metastable \ce{^{99m}}Tc isotopes, which are used extensively in medical science, are produced by MuCIP on \ce{^{Nat}}Mo and $^{100}$Mo.
\end{abstract}

\end{frontmatter}


\section{Introduction}

Radioactive isotopes (RIs) for science and technology are produced by various kinds of nuclear reactions on stable nuclei.
So far, neutron-induced reactions have been extensively used for RI production because of the large neutron flux available at nuclear reactors.
The typical reaction used is the neutron capture (n,$\gamma$) reaction. 
Recently, the photon capture reaction has been shown to be very effective for selectively producing RIs by using the ($\gamma$,xn) reaction with $x$ being 1 or 2, depending on the photon energy \cite{eji11,szp13}. 
Here the cross section is large in the E1 giant resonance region, and high-flux photons are available. 
Recently, Compton back-scattering of laser photons scattered off GeV electrons in a storage ring has been shown to be effective to produce high-density RIs\cite{eji11}. 
RIs produced from the target isotope $^A_Z$X are mainly $^{A+1}_{Z}$X and $^{A-1}_Z$X with $A$ and $Z$ being the mass and atomic numbers in the neutron capture and photon capture reactions, respectively.
They are RIs with the different mass number but the same atomic number as the target isotope. 

Muon capture isotope production (MuCIP) by the ordinary muon capture method (OMC) is used to produce RIs of $^{A-x}_{Z-1}$X with atomic numbers of 1 less by than the target isotope.
Furthermore, the $\mu$ capture probability (cross section) is large enough to stop in light and heavy nuclei using the recently developed high intensity $\mu$ beams.
In previous works we studied isotope distributions of residual nuclei after OMC in order to investigate weak nuclear responses associated with double beta decays \cite{eji05,eji10,eji16,has16}. 
We have also shown that the $\mu$ capture isotope detection (MuCID) can be used to detect nuclear isotopes by measuring $\gamma$-rays following OMC \cite{eji13}.
These works suggest that OMC can be used to produce RIs as well.
OMCs have been studied by measuring protons, neutrons, charged particles and $\gamma$ rays following the $\mu$ capture into the nucleus, as given in this review article and references therein \cite{mae01,luc73,eva73,mae07_1,mae07_2,mae07_3}.

The present work aims to show that MuCIP using OMC is useful at producing new types of RIs and to demonstrate the feasibility of MuCIP by using OMC on Mo isotopes.
In this work, we have extended the neutron cascade model for neutron emission following OMC to the neutron proton emission model by including a possible proton emission. 
Section 2 describes the OMC RI production and the proton neutron emission model. 
The OMC RI productions for several medium-to-heavy nuclei are presented in section 3. 
The Nb RI production rates for OMC on natural Mo isotopes are reported in section 4. 
Section 5 gives several remarks on MuCIP.

\begin{figure}[tbp]
\begin{center}
\includegraphics[width=0.7 \columnwidth]{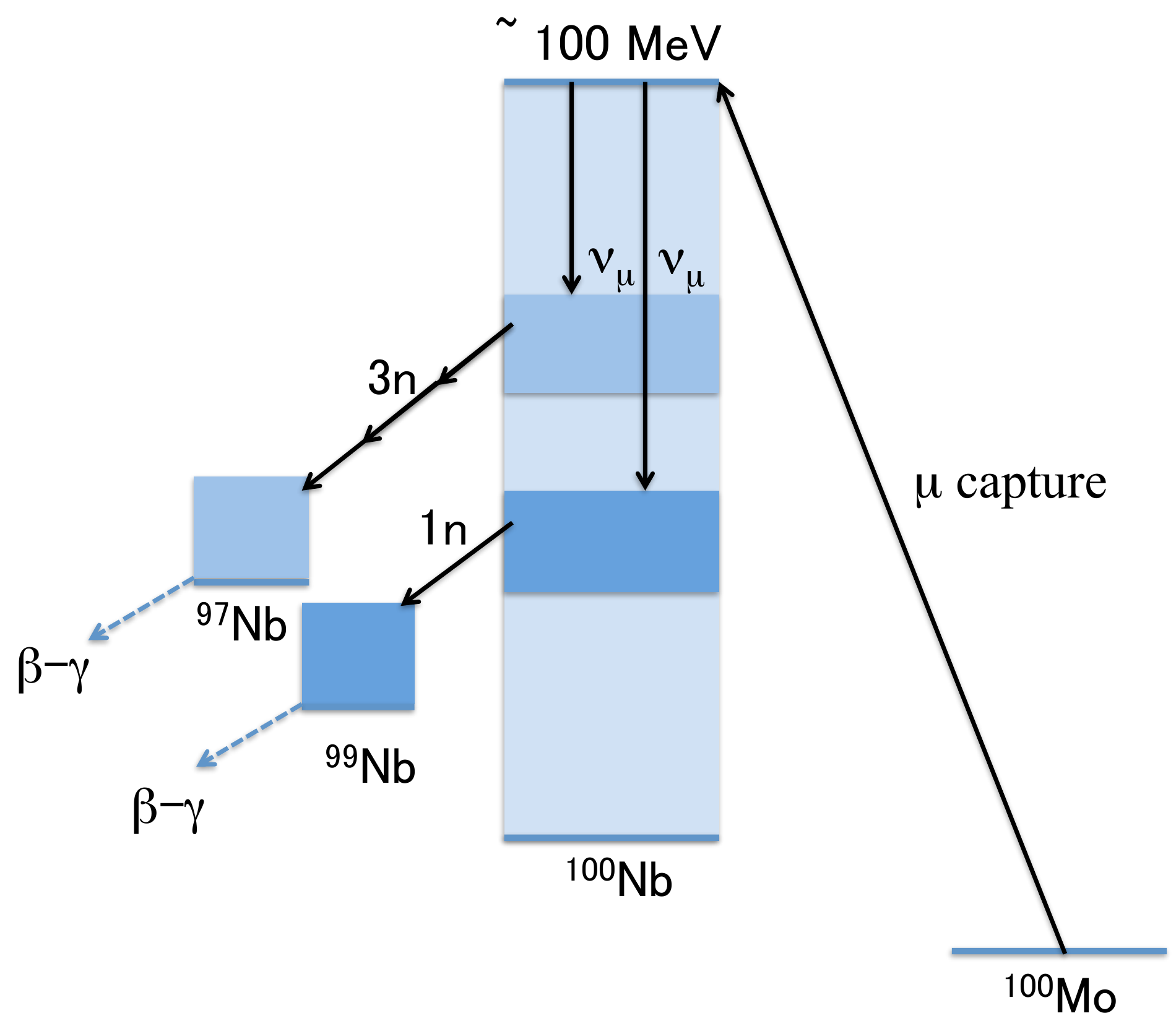}
\end{center}
\caption{Nuclear excitation and de-excitation scheme associated with OMC on $^{100}$Mo. Highly excited states with $E\approx$ 30 MeV likely de-excite by emitting $x\approx $3 neutrons, resulting in production of the $^{97}$Nb RI, while those with $E\approx$ 15 MeV de-excite by $x\approx$ 1 neutron emission, resulting in the $^{99}$Nb RI. The residual RIs are identified by measuring $\beta$-delayed $\gamma$ rays characteristic of the residual nucleus X. $B(\mu,$E) is the $\mu$ capture strength (response) with $E$ being the excitation energy.
\label{fig:1}}
\end{figure}

\section{Radioactive isotope production by OMC}

MuCIP has several unique features for RI production. 
Low-energy negative muons are stopped in the target atoms and are exclusively trapped at the inner atomic orbits for about a few nanoseconds. 
Then, after some 100 nanoseconds, in the case of medium-to-heavy nuclei, most of them are captured into the nucleus and some decay via the weak interaction through the emission of an electron, an electron anti-neutrinos ($\overset{-}{\nu_e}$) and a muon neutrinos ($\nu_{\mu}$).
The $\mu$ capture probability is more than 90$\%$ for most nuclei with $Z\geq$20\cite{nag03}. 
The negative $\mu$ capture into the target nucleus of $^A_Z$X is followed mostly by a number ($x$) of neutron emission and $y$ proton emission for medium-heavy nuclei but the number of charge particle emission including proton and alpha is greater in light nuclei\cite{mae01,mae07_1,mae07_2,mae07_3}. 
Then the residual RIs are $^{A-x}_{Z-1}$X with the atomic number $Z-1$.

Several nuclei of $^{A-x}_{Z-1}$X with $x$ = 0, 1, 2, 3, 4, 5 are produced.
Some of them are likely to be radioactive nuclei within the appropriate half-life and $\gamma$-ray energy. 
In other words, one will always end up with some RIs with the atomic number $Z-1$ from the target nuclei with the atomic number $Z$.
The location of the RIs in the target is defined by the $\mu$ beam spot and the depth where the $\mu$ stops.
Here the $\mu$ momentum (energy) and intensity is adjusted by the bending magnet of the beamline.
For target with thickness less than 1 mm, the use of energy degrader could help to reduce the momentum intensity and enhanced muon stopping within the target, however this method will increase the momentum spread.
A high-efficiency beam channel like MuSIC at RCNP, Osaka University makes it realistic to attain a high flux of muons of the order of 10$^{4}$ - 10$^{5}$ per second by using a high-intensity ($\mu$A - mA) medium energy proton beam \cite{Cook17}.

The $\mu$ capture on a target nucleus is expressed as the weak charge exchange process $^{A}_{Z}$X + $\mu \rightarrow ^{A}_{Z-1}$X$^*$ +$\nu_{\mu}$. 
The $\mu$ brings into the target nucleus at an excitation energy of 100 MeV around the $\mu$ mass but almost no momentum, and the $\nu_{\mu}$ takes away most of the energy around 95-50 MeV, leaving the residual excited nucleus of $^A_{Z-1}$X* with an excitation energy of around 5-50 MeV. 
It de-excites mainly by emitting neutrons in the case of medium-to-heavy nucleus \cite{mae01,luc73,eva73}. 
Among these reaction channels, the major reaction channel is one neutron emission. 

MuCIP on a medium-to-heavy nucleus proceeds mainly by the OMC neutron emission. It is expressed as  
\begin{equation}
^{A}_{Z}X + \mu  \rightarrow  ^{A-x}_{Z-1}X +\nu_{\mu}+xn + \gamma, 
\end{equation}
where the number $x$ of the emitted neutron is $x$ = 0-5, depending on the excitation energy. 
The residual nucleus $^{A-x}_{Z-1}$X = X' may be radioactive, also depending on the mass number $A-x$. 
The RI production rate $R(X')$ per sec and the number $N(X')$ of RIs X' by the $\mu$ beam exposure for $T_{\mu}$ sec are given as 
\begin{equation}
R(X')=\eta R(\mu X) Br(X') N_{\mu},~~~
N(X')=R(X') k(T_{\mu}),
\end{equation}
where $\eta$ is the $\mu$ stopping probability, $R(\mu X)$ is the $\mu$ capture probability, $Br(X')$ is the branching ratio to the isotope X' channel via $x$ neutron emission and $N_{\mu}$ is the $\mu$ flux and $k(T{\mu}$) is the RI decay rate during the muon beam exposure.
By using a thick target and adjusting the $\mu$ momentum so as to stop in the target, one can get $\eta \approx $1.
The muons are then mostly captured into the nucleus in the case of medium-to-heavy nuclei with $Z\geq$20. 
The branching ratio is evaluated by using the pre-equilibrium (PEQ) - equilibrium (EQ) proton neutron-emission model \cite{eji89}. Reference \cite{has17} shows that the systematic study of PEQ to EQ ratio is fixed to 25$\%$ to reproduce neutron spectra in experimental observation.

The excited state with the excitation energy $E_{ex}$ in the $^A_{Z-1}$X after the ($\mu,\nu_{\mu}$) capture reaction de-excites by emitting neutrons at the PEQ and EQ stages \cite{eji89}.
First, we consider the major decay process in the neutron emission model \cite{has15,has18}. The neutron decay scheme is illustrated in Fig. 1.
The excited state produced by the capture decays by emitting neutrons if the state is neutron-unbound, and decays by emitting $\gamma$ rays to the ground state if it is bound. 
Here we follow the neutron emission model(NEM) in references \cite{has15} and \cite{has18}.
In previous work \cite{has15,has18}, we ignored the proton emission which is strongly suppressed by the Coulomb barrier in the case of medium and heavy nuclei. 
The energy spectrum of the first neutron n(1) is given by \cite{eji89}
\begin{equation}
S(E_{n(1)}) = k [E_{n(1)}exp (-\frac{E_{n(1)}}{T_{EQ}(E)}) + p E_{n(1)}exp (-\frac{E_{n(1)}}{T_{PEQ}(E)})],
\end{equation}
where $T_{EQ}(E)$ and $T_{PEQ}(E)$ are the EQ and PEQ nuclear temperatures and $p$ is the fraction of the PEQ neutron emission.
We used the values of $T_{EQ}(E)=\surd(E_{ex}/a)$ with $a$ as the level density parameter \cite{eji89}. 
The parameter is expressed as $a$ = $A$/8 MeV for the nucleus with the mass number $A$. 
$T_{PEQ}(E)$ is given by $b \times T_{EQ}(E)$ with $b\approx$3 for the OMC medium-excitation ($E_{ex}\approx$ 10-40 MeV). 
After one neutron emission, the emission takes places only via the EQ.
Thus, the neutron emission follows only the first term of eq (3).

The residual nucleus $^{A-1}_{Z-1}$X after the first neutron emission de-excites by emitting a second neutron or $\gamma$ rays, depending on the excitation energy above or below the neutron threshold energy, $B_{n}$. 
The ground state of $^{A-1}_{Z-1}$X is populated after the $\gamma$ decay. 
The second neutron n(2) is the EQ evaporation neutron, and then if the residual nucleus after the second neutron emission is neutron-unbound, the third neutron is emitted, and the cascade neutron emission continues until the residual state becomes neutron-bound, which then decays by emitting gamma rays.
Finally, the residual isotopes of $^{A-x}_{Z-1}$X with $x$ = 0, 1, 2, 3, ... , are attained, depending on the initial excitation energy $E_{ex}$ and the number $x$ of the emitted neutrons.
Some of the residual isotopes of $^{A-x}_{Z-1}$X produced by MuCIP are $\beta$-unstable RIs. 

In the previous NEM \cite{has17,has15,has18}, the mass number $A-x$ distribution is shown to reflect the strength distribution $B(\mu,E)$ of the nucleus $^A_{Z-1}$X after the OMC. 
The giant resonance(GR)-like broad peak at around 10-15 MeV corresponds to the excitation region of one neutron emission. 
Therefore, the one neutron-emission process gets dominant.
The RIs population of $^{A-1}_{Z-1}$X after 1 neutron emission, the population of $^{A-x}_{Z-1}$X decrease as $x$ increases beyond $x$=1. 

Now let us extend the NEM to the proton neutron emission model(PNEM), where a proton is emitted in a special case as given in Fig. 3.
If the $E_{ex}$ is above both $B_n$ and $B_p$, it decays by emitting predominantly neutrons since proton emission is prohibited by the Coulomb barrier in the case of medium and heavy nuclei. 
Decays by proton emission is limited to the region of $B_n$ $\geq E \geq$ $B_p$.
Note that after a proton emission, the residual state decays by emitting $\gamma$ rays to the ground state. 

\begin{figure}[tbp]
\begin{center}
\includegraphics[width=0.7\columnwidth]{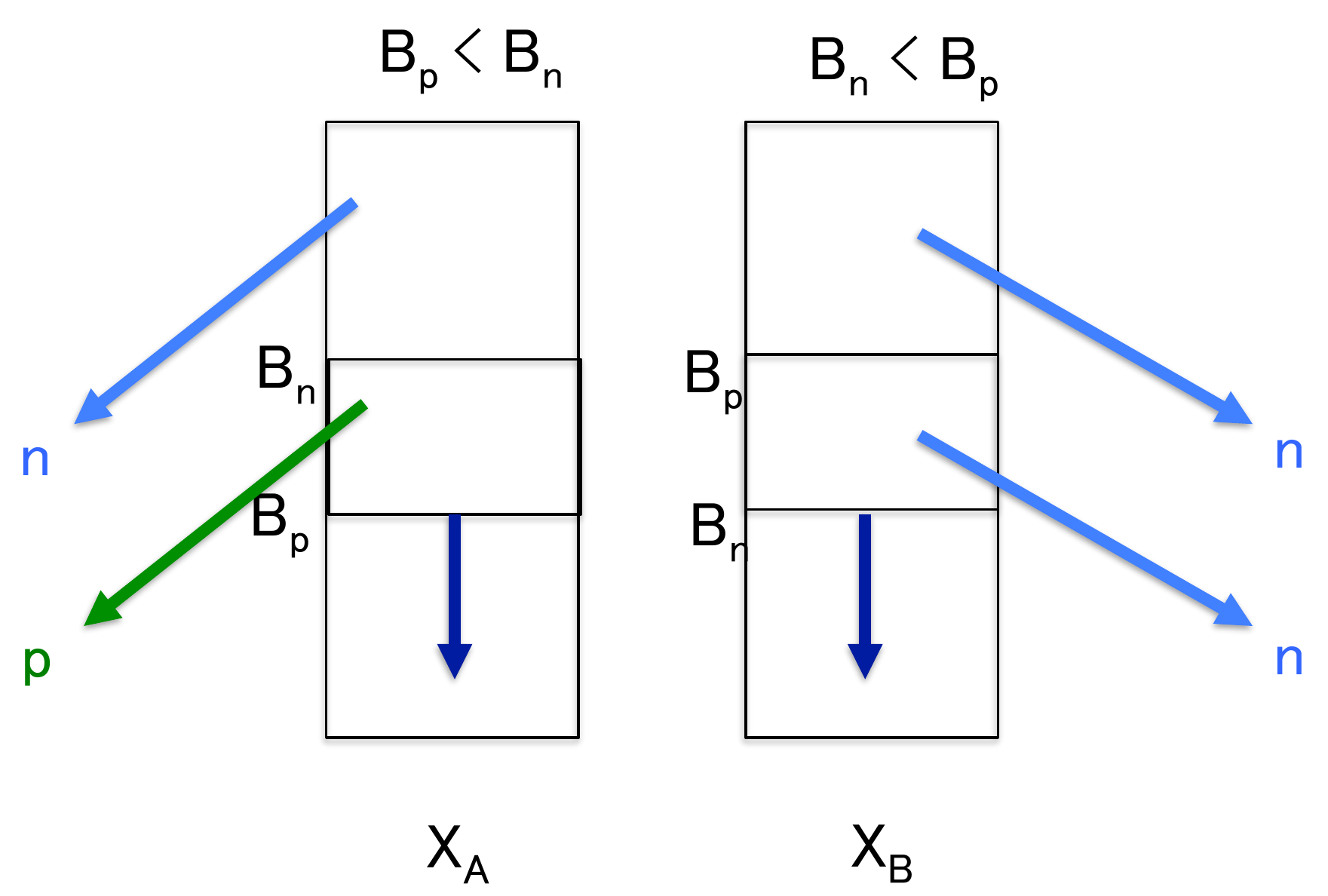}
\end{center}
\caption{PEQ-EQ proton and neutron emission schemes in the cases of X$_A$ with $B_n$ $\geq E_{ex} \geq$ $B_p$ and X$_B$ with $B_p$ $\geq E_{ex} \geq$ $B_n$ with $B_n$, $E_{ex}$ and $B_p$ are the neutron binding energy, the excitation energy and the proton binding energy.
\label{fig:2}}
\end{figure}

\section{OMC GR region during RI production} 

It is shown that the OMC preferentially excites the GR region with $E\approx$ 10-40 MeV in the nucleus $^{A}_{Z-1}$X \cite{eji16,has15,has18,rap67}. 
Then the strength distribution of $B(\mu,E)$ is given by the sum of the 2 giant resonance strengths of $B_1(\mu,E) $ and $B_2(\mu,E)$ \cite{has15,has18},

\begin{equation}
B(\mu,E)=B_1(\mu,E) + B_2(\mu,E),  
\end{equation}

\begin{equation}
B_i(\mu,E)=\frac{B_i(\mu)}{(E-E_{Gi})^2 + (\Gamma_{i}/2)^2},
\end{equation}
where $E_{Gi}$ and $\Gamma_i$ with $i$=1,2 are the resonance energy and the width for the $i$th giant resonance, and the constant $B_i(\mu)$ is expressed as $B_i(\mu)=\sigma_i \Gamma_i/(2\pi)$ with $\sigma_i$ being the total strength integrated over the excitation energy.

RI mass distributions for target isotopes of $^{100}$Mo, $^{107}$Pd, $^{108}$Pd, $^{127}$I, and $^{209}$Bi, as typical examples are evaluated by means of the PEQ and EQ proton and neutron model as shown in Fig. 3.  
The obtained RI mass distributions compared with previously observed experimental data in Tables 1-5. 

\begin{table}[tbp]
\caption{RIs produced by OMC on $^{100}$Mo.
Columns 1 and 2 show the reaction and RI produced by $\mu$-capture. Columns 3 and 4 gives the experimental and current calculation.
\label{tab:1}} 
\begin{center}
\small\addtolength{\tabcolsep}{-5pt}
\begin{tabular}{|c|c|c|c|}
\hline
Reaction & Final N & ~~Experimental~~ & ~~This model~~  \\
\hline\hline
$^{100}$Mo($\mu$,0n) & $^{100}$Nb & 8 \cite{has18} & 12$\pm$4 \\
$^{100}$Mo($\mu$,1n) & $^{99}$Nb & 51 \cite{has18} & 58$\pm$7 \\
$^{100}$Mo($\mu$,2n) & $^{98}$Nb & 16 \cite{has18} & 15$\pm$1 \\
$^{100}$Mo($\mu$,3n) & $^{97}$Nb & 13 \cite{has18} & 9$\pm$4  \\
$^{100}$Mo($\mu$,4n) & $^{96}$Nb & 6 \cite{has18} & 4$\pm$2 \\
$^{100}$Mo($\mu$,5n) & $^{95}$Nb & 3 \cite{has18} & 2$\pm$1 \\
$^{100}$Mo($\mu$,1p) & \ce{^{99m}}Tc & 0 \cite{has18} & 0 \\
$^{100}$Mo($\mu$,1n1p) & $^{98}$Tc & 0 \cite{has18} & 0 \\
$^{100}$Mo($\mu$,2n1p) & $^{97}$Tc & 0 \cite{has18} & 0  \\
$^{100}$Mo($\mu$,3n1p) & $^{96}$Tc & 0 \cite{has18} & 0 \\
 & Total, $\Sigma$($\%$) & 97 & 100 \\
\hline
\end{tabular}
\end{center}
\end{table}
 
\begin{table}[tbp]
\caption{RIs produced by OMC on $^{107}$Pd. 
Columns 1 and 2 show the reaction and RI produced by $\mu$-capture. Columns 3 and 4 gives the experimental and current calculation.
\label{tab:2}} 
\begin{center}
\small\addtolength{\tabcolsep}{-5pt}
\begin{tabular}{|c|c|c|c|}
\hline
Reaction & Final N & Experimental & This model  \\
\hline\hline
$^{107}$Pd($\mu$,0n) & $^{107}$Rh & - & 13.3$\pm$3.7  \\
$^{107}$Pd($\mu$,1n) & $^{106}$Rh & - & 55.6$\pm$10.7 \\
$^{107}$Pd($\mu$,2n) & $^{105}$Rh & - & 15.0$\pm$2.4 \\
$^{107}$Pd($\mu$,3n) & $^{104}$Rh & - & 6.0$\pm$4.6  \\
$^{107}$Pd($\mu$,4n) & $^{103}$Rh & - & 1.6$\pm$3.0 \\
$^{107}$Pd($\mu$,5n) & $^{102}$Rh & - & 0.6$\pm$2.2 \\
$^{107}$Pd($\mu$,6n) & $^{101}$Rh & - & 0.18$\pm$2.1 \\
$^{107}$Pd($\mu$,7n) & $^{100}$Rh & - & 0.0019$\pm$2.4 \\
$^{107}$Pd($\mu$,1p) & $^{106}$Ru & - & 3.9$\pm$2 \\
$^{107}$Pd($\mu$,1n1p) & $^{105}$Ru & - & 0 \\
$^{107}$Pd($\mu$,2n1p) & $^{104}$Ru & - & 3.0$\pm$2  \\
$^{107}$Pd($\mu$,3n1p) & $^{103}$Ru & - & 0.2$\pm$1 \\
$^{107}$Pd($\mu$,4n1p) & $^{102}$Ru & - & 0.4$\pm$2 \\
$^{107}$Pd($\mu$,5n1p) & $^{101}$Ru & - & 0.08$\pm$1 \\
$^{107}$Pd($\mu$,6n1p) & $^{100}$Ru & - & 0.03$\pm$1 \\
 & Total, $\Sigma$($\%$) & - & 99.9 \\
\hline
\end{tabular}
\end{center}
\end{table}

\begin{table}[tbp]
\caption{RIs produced by OMC on $^{108}$Pd. 
Columns 1 and 2 show the reaction and RI produced by $\mu$-capture. Columns 3 and 4 gives the experimental and current calculation.
\label{tab:3}} 
\begin{center}
\small\addtolength{\tabcolsep}{-5pt}
\begin{tabular}{|c|c|c|c|}
\hline
Reaction & Final N & Experimental & This model  \\
\hline\hline
$^{108}$Pd($\mu$,0n) & $^{108}$Rh & - & 9.9$\pm$3.8  \\
$^{108}$Pd($\mu$,1n) & $^{107}$Rh & - & 58.8$\pm$10.7 \\
$^{108}$Pd($\mu$,2n) & $^{106}$Rh & - & 14.9$\pm$2.4 \\
$^{108}$Pd($\mu$,3n) & $^{105}$Rh & - & 8.3$\pm$4.6  \\
$^{108}$Pd($\mu$,4n) & $^{104}$Rh & - & 1.8$\pm$3.1 \\
$^{108}$Pd($\mu$,5n) & $^{103}$Rh & - & 0.8$\pm$2.2 \\
$^{108}$Pd($\mu$,6n) & $^{102}$Rh & - & 0.2$\pm$2.1 \\
$^{108}$Pd($\mu$,7n) & $^{101}$Rh & - & 0.013$\pm$2.4 \\
$^{108}$Pd($\mu$,8n) & $^{100}$Rh & - & 0 \\
$^{108}$Pd($\mu$,1p) & $^{107}$Ru & - & 0 \\
$^{108}$Pd($\mu$,1n1p) & $^{106}$Ru & - & 3.9$\pm$2 \\
$^{108}$Pd($\mu$,2n1p) & $^{105}$Ru & - & 0  \\
$^{108}$Pd($\mu$,3n1p) & $^{104}$Ru & - & 1.0$\pm$1 \\
$^{108}$Pd($\mu$,4n1p) & $^{103}$Ru & - & 0.07$\pm$2 \\
$^{108}$Pd($\mu$,5n1p) & $^{102}$Ru & - & 0.22$\pm$3 \\
$^{108}$Pd($\mu$,6n1p) & $^{101}$Ru & - & 0.017$\pm$1 \\
$^{108}$Pd($\mu$,7n1p) & $^{100}$Ru & - & 0.00027$\pm$1 \\
 & Total, $\Sigma$($\%$) & - & 99.9 \\
\hline
\end{tabular}
\end{center}
\end{table}

\begin{table}[tbp]
\caption{RIs produced by OMC on $^{127}$I. 
Columns 1 and 2 show the reaction and RI produced by $\mu$-capture. Columns 3 and 4 gives the experimental and current calculation.
\label{tab:4}} 
\begin{center}
\small\addtolength{\tabcolsep}{-5pt}
\begin{tabular}{|c|c|c|c|}
\hline
Reaction & Final N & Experimental & This model  \\
\hline\hline
$^{127}$I($\mu$,0n) & $^{127}$Te & 7$\pm$3 \cite{mae07_3}  & 7.0$\pm$3  \\
$^{127}$I($\mu$,1n) & $^{126}$Te & 44$\pm$3 \cite{mae07_3}  & 54.6$\pm$7 \\
$^{127}$I($\mu$,2n) & $^{125}$Te & 15$\pm$3 \cite{mae07_3}  & 21.4$\pm$5 \\
$^{127}$I($\mu$,3n) & $^{124}$Te & 15$\pm$2 \cite{mae07_3}  & 10.7$\pm$3  \\
$^{127}$I($\mu$,4n) & $^{123}$Te & 8$\pm$5 \cite{mae07_3}  & 2.7$\pm$1 \\
$^{127}$I($\mu$,5n) & $^{122}$Te & 1.5$\pm$10 \cite{mae07_3}  & 1.2$\pm$10 \\
$^{127}$I($\mu$,6n) & $^{121}$Te & 1.0$\pm$5 \cite{mae07_3}  & 0.6$\pm$5 \\
$^{127}$I($\mu$,7n) & $^{120}$Te & 0 \cite{mae07_3}  & 0.2$\pm$5 \\
$^{127}$I($\mu$,1p) & $^{126}$Sb & 0 \cite{mae07_3}  & 0 \\
$^{127}$I($\mu$,1n1p) & $^{125}$Sb & 0 \cite{mae07_3}  & 1.7$\pm$5 \\
$^{127}$I($\mu$,2n1p) & $^{124}$Sb & 0 \cite{mae07_3}  & 0  \\
$^{127}$I($\mu$,3n1p) & $^{123}$Sb & 0 \cite{mae07_3}  & 0.45$\pm$5 \\
$^{127}$I($\mu$,4n1p) & $^{122}$Sb & 0 \cite{mae07_3}  & 0 \\
$^{127}$I($\mu$,5n1p) & $^{121}$Sb & 0 \cite{mae07_3}  & 0.12$\pm$5 \\
$^{127}$I($\mu$,6n1p) & $^{120}$Sb & 0 \cite{mae07_3}  & 0.0029$\pm$1 \\
 & Total, $\Sigma$($\%$) & 97.3 & 99.9 \\
 \hline
\end{tabular}
\end{center}
\end{table}

\begin{table}[tbp]
\caption{RIs produced by OMC on $^{209}$Bi. 
Columns 1 and 2 show the reaction and RI produced by $\mu$-capture. Columns 3 and 4 gives the experimental and current calculation.
\label{tab:5}} 
\begin{center}
\small\addtolength{\tabcolsep}{-5pt}
\begin{tabular}{|c|c|c|c|}
\hline
Reaction & Final N & Experimental & This model  \\
\hline\hline
$^{209}$Bi($\mu$,0n) & $^{209}$Pb & 3$\pm$1 \cite{mae07_3}  & 7$\pm$3  \\
$^{209}$Bi($\mu$,1n) & $^{208}$Pb & 46$\pm$4 \cite{mae07_3}  & 54.6$\pm$7 \\
$^{209}$Bi($\mu$,2n) & $^{207}$Pb & 30$\pm$3 \cite{mae07_3}  & 21.4$\pm$5 \\
$^{209}$Bi($\mu$,3n) & $^{206}$Pb & 10$\pm$3 \cite{mae07_3}  & 10.7$\pm$3  \\
$^{209}$Bi($\mu$,4n) & $^{205}$Pb & 5$\pm$1 \cite{mae07_3}  & 2.7$\pm$2 \\
$^{209}$Bi($\mu$,5n) & $^{204}$Pb & 1$\pm$1 \cite{mae07_3}  & 1.2$\pm$1 \\
$^{209}$Bi($\mu$,6n) & $^{203}$Pb & 0 \cite{mae07_3}  & 0.59$\pm$5 \\
$^{209}$Bi($\mu$,7n) & $^{202}$Pb & 0 \cite{mae07_3}  & 0.24$\pm$5 \\
$^{209}$Bi($\mu$,8n) & $^{201}$Pb & 0 \cite{mae07_3}  & 0.008$\pm$5 \\
$^{209}$Bi($\mu$,1p) & $^{208}$Tl & 0 \cite{mae07_3}  & 0 \\
$^{209}$Bi($\mu$,1n1p) & $^{207}$Tl & 0 \cite{mae07_3}  & 0 \\
$^{209}$Bi($\mu$,2n1p) & $^{206}$Tl & 0 \cite{mae07_3}  & 0  \\
$^{209}$Bi($\mu$,3n1p) & $^{205}$Tl & 0 \cite{mae07_3}  & 0.98$\pm$5 \\
$^{209}$Bi($\mu$,4n1p) & $^{204}$Tl & 0 \cite{mae07_3}  & 0.13$\pm$5 \\
$^{209}$Bi($\mu$,5n1p) & $^{203}$Tl & 0 \cite{mae07_3}  & 0.25$\pm$6 \\
$^{209}$Bi($\mu$,6n1p) & $^{202}$Tl & 0 \cite{mae07_3}  & 0.079$\pm$5 \\
$^{209}$Bi($\mu$,7n1p) & $^{201}$Tl & 0 \cite{mae07_3}  & 0.035$\pm$5 \\
 & Total, $\Sigma$($\%$) & 95.2 & 99.9 \\
 \hline
\end{tabular}
\end{center}
\end{table}

The relative populations of the isotopes with $A-1$ is around 50-60 $\%$, while those with $A, A-2, A-3$ are around 5-20 $\%$. 
The proton emission channel is indeed small, being at most a few $\%$ in the present nuclei.

\begin{figure}[tbp]
\begin{center}
\includegraphics[width=1.0\textwidth]{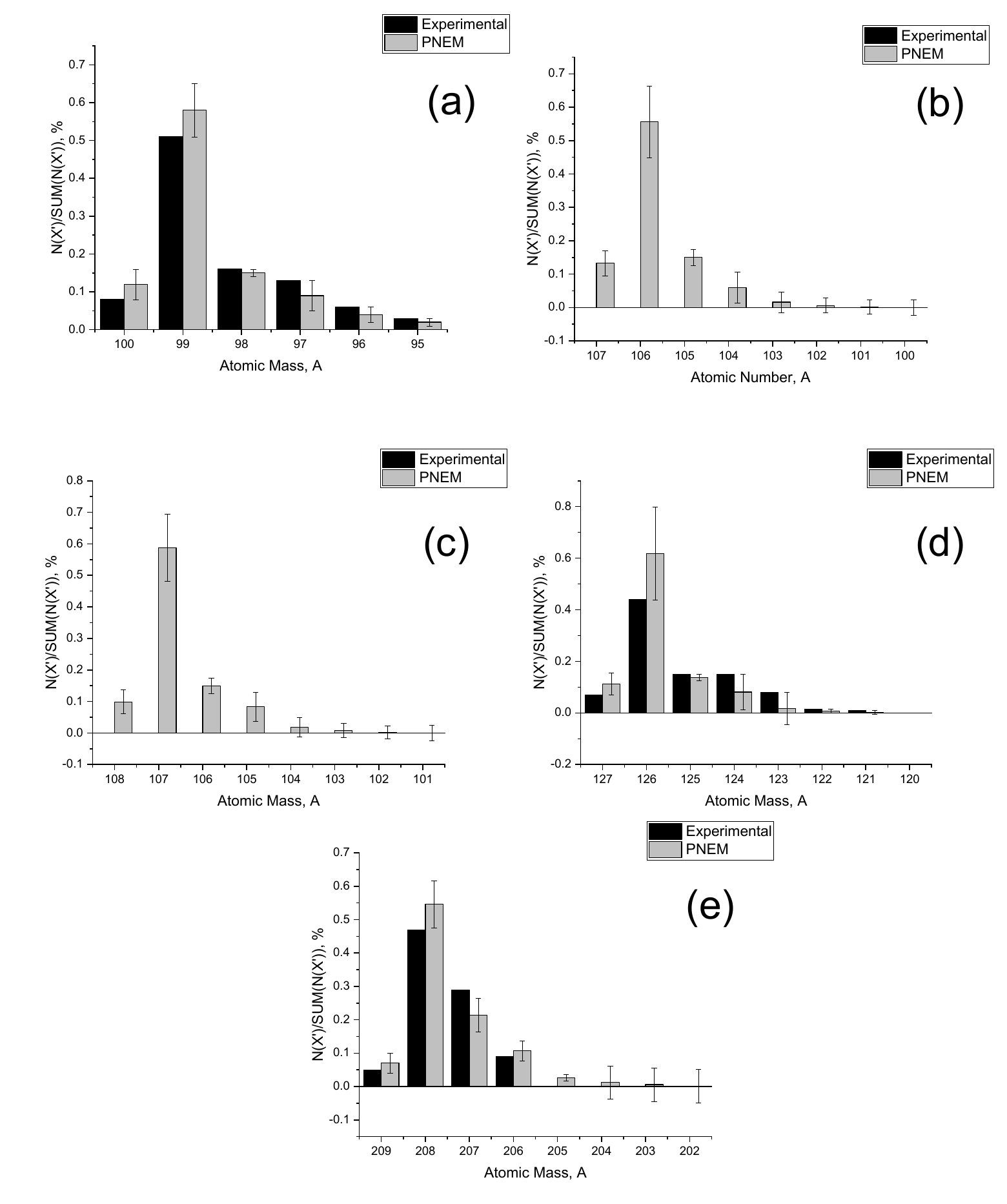}
\end{center}
\caption{RI mass distribution from experimental observations and calculations for neutron and proton emission events after muon capture on (a) Molybdenum-100, (b) Palladium-107, (c) Palladium-108 (d) Iodine-127 and (e) Bismuth-209.  
\label{fig:3}}
\end{figure}

From the comparison, the pattern of the GR$_1$ excitation region corresponds to the energy window for the 1 neutron emission since the energy, $E_{G1}\approx $10-18 MeV, is just above the 1 neutron threshold (binding) energy and below the 2 neutrons threshold energy. 
The GR$_2$ excitation region corresponds to the energy window for the 3 and 4 neutron emission populations at $\approx$ 25-45 MeV.
The parameters of $E_{G1}$ and $E_{G2}$ as a function of A are concluded as $E_{G1}$ = 30$\times$ A$^{-1/5}$ MeV and $E_{G2}$ = 75 $\times$ A$^{-1/5}$ MeV.

\section{RI productions by OMC on \ce{^{Nat}}Mo} 

Muon capture reactions on natural Mo target were measured to demonstrate the feasibility of MuCIP at the MuSIC beamline at RCNP, Osaka University. 
The negative pions were produced by the 400 MeV proton beam with 1 $\mu$A from the RCNP ring cyclotron. 
Then, the MuSIC transports the low-momentum negative muons to the beam exit where the OMC target was placed about 10 cm from it. 
The muons were stopped at the natural Mo target made of 4 Mo plates, each with 0.5$\times $5$\times $5 cm$^3$. 
The isotopic abundance ratios of the target were $^{92}$Mo 0.158, $^{94}$Mo 0.091, $^{95}$Mo 0.157, $^{96}$Mo 0.165, $^{97}$Mo 0.095, $^{98}$Mo 0.238 and $^{100}$Mo 0.096.  
In 2012, the $\mu$-beam spot was very large compared to the small Mo target. 
The target was approximately 6 $\%$ of the total $\mu$-beam spot. 
The $\mu$ intensity expected to stop at the target was about 3$\times$10$^6$ per sec.
Note that the muon beam spot have been improved much in the new beam line \cite{Cook17}.
 
\begin{figure}[tbp]
\begin{center}
\includegraphics[width=1.0\textwidth]{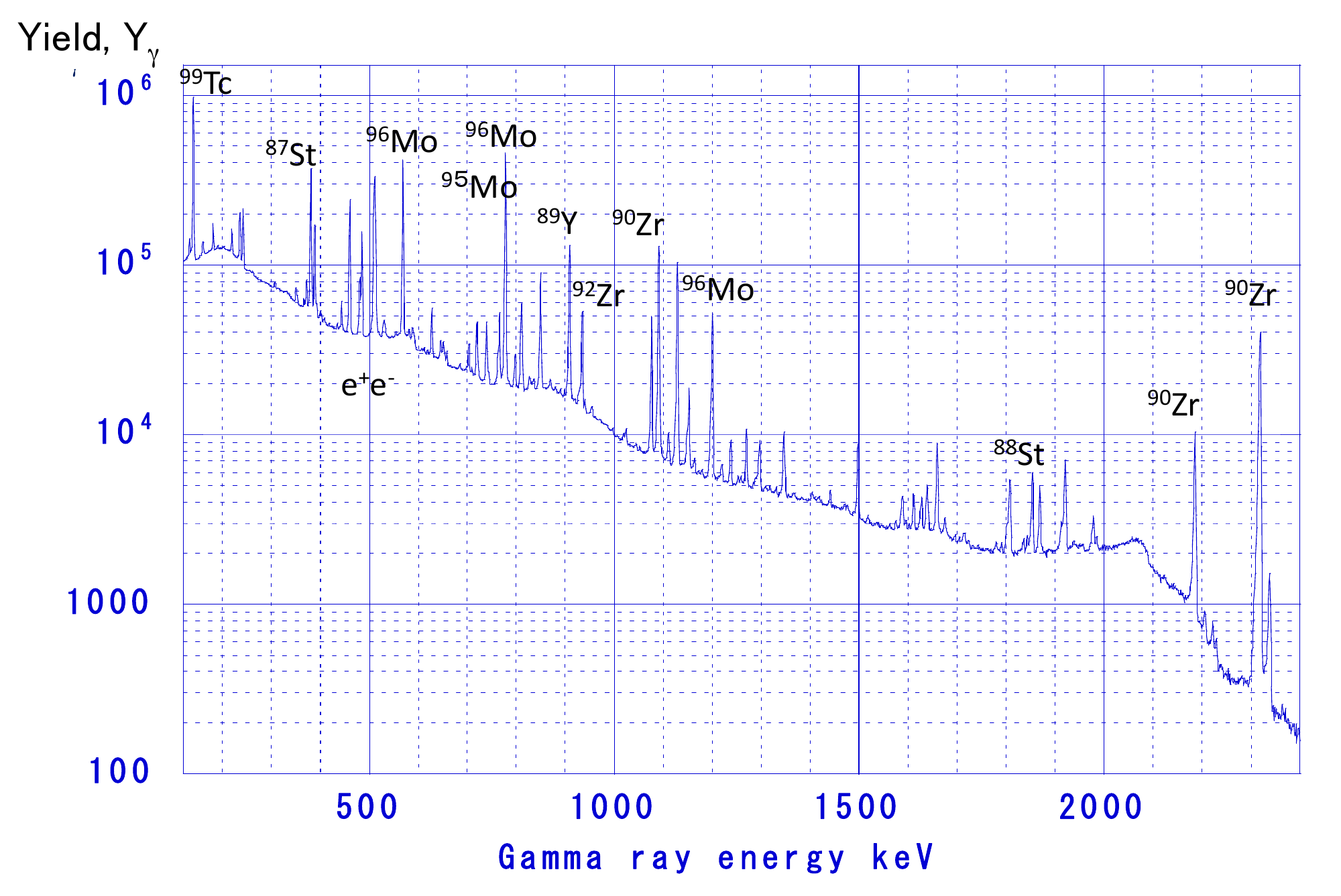}
\end{center}
\caption{Energy spectrum of delayed $\gamma$ rays from long-lived RIs produced by OMC on \ce{^{Nat}}Mo. The spectrum was measured 22.25 hours after $\mu$ irradiation. 
\label{fig:4}}
\end{figure}

\begin{landscape}
\begin{table}[tbp]
\caption{RIs produced by OMC on \ce{^{Nat}}Mo. 
Columns 1 and 2 show the RI produced by$\mu$ capture 
and the residual nucleus, and column 3 gives the emission process involved.
Column 4 gives the half-life of the RIs produced by OMC. 
Column 5 is the number of the RIs, column 6 lists the typical $\gamma$ ray(s)\cite{eji13}, and column 7 is the calculated N(X') by the PNEM.}
\label{tab:6}
\centering
\begin{threeparttable}
\begin{tabular}{|c|c|c|c|c|c|c|}
\hline
RI         & Final N    & Process                       & Half-life (hr)                & N(X')$\times$10$^8$ & $\gamma$ rays (keV)   & calc. N(X')$\times$10$^8$                  \\ 
\hline
$^{100}$Nb & $^{100}$Mo & $^{100}$Mo($\mu$,0n)          & 4.4 $\times$ 10$^{-3}$        & 0.6 $\pm$ 0.1      & 535.6\tnote{a}        & 0.35$\pm$0.26 \\
$^{99}$Mo  & \ce{^{99m}}Tc  & $^{100}$Mo($\mu$,n$\beta ^-$) & 66                            & 3.8 $\pm $0.4       & 140.5, 181.0, 739.5   & 2.91$\pm$1.02                   \\
$^{98}$Nb  & $^{98}$Mo  & $^{100}$Mo($\mu$,2n)          & 7.1 $\times$ 10$^{-3}$, 0.855 & 3.0 $\pm$0.8        & 734.7\tnote{a} , 787.4 & 2.08$\pm$1.01 \\
$^{97}$Nb  & $^{97}$Mo  & $^{98}$Mo($\mu$,1n)           & 1.2                           & 8.8 $\pm$1.5        & 658.1   & 8.51$\pm$0.83                                \\
$^{97}$Zr  & $^{97}$Nb  & $^{98}$Mo($\mu$,p)            & 16.9                          & 0.05 $\pm $0.02       & 743.5   &   --                              \\
$^{96}$Nb  & $^{96}$Mo  & $^{97}$Mo ($\mu$,1n)          & 23.4                          & 4.5 $\pm$ 1.0      & 568.8, 778.2, 1091.3    & 7.02$\pm$1.37                \\
$^{95}$Nb  & $^{95}$Mo  & $^{96}$Mo($\mu$,1n)           & 1205                          & 6.7 $\pm $1.0       & 765.8           & 7.52$\pm$2.16                          \\
$^{94}$Nb  & $^{94}$Mo  & $^{98}$Mo($\mu$,1n)           & 1.75 $\times$ 10$^8$          & 8.62 $\pm $1.0\tnote{b}                   & --              & 8.29$\pm$1.13                        \\
$^{93}$Nb  & $^{93}$Nb  & $^{94}$Mo($\mu$,1n)           & 1.41 $\times$ 10$^5$          & 5.26$\pm $1.0\tnote{b}                   & --              & 5.06$\pm$1.35                        \\
$^{92}$Nb  & $^{92}$Zr  & $^{94}$Mo ($\mu$,2n)          & 244.8                         & 3.0 $\pm$ 0.15      & 934.5           & 2.78$\pm$1.17                        \\
$^{91}$Nb  & $^{91}$Zr  & $^{92}$Mo ($\mu$,1n)          & 6 $\times$ 10$^6$             & 5.19$\pm $1.0\tnote{b}                  & --              & 5.00$\pm$1.17                        \\
$^{90}$Nb  & $^{90}$Zr  & $^{92}$Mo ($\mu$,2n)          & 14.6                          & 1.9 $\pm $0.3       & 1129.2, 2186, 2319.0  & --                  \\
\hline
\end{tabular}
\begin{tablenotes}
    \item[a] The $\gamma$ rays measured in the $^{100}$Mo experiment.
    \item[b] N(X') obtained by calculation using PNEM.
\end{tablenotes}
\end{threeparttable}
\end{table}
\end{landscape}

Delayed $\gamma$ rays from RIs produced in the 4 Mo plates (A, B, C, D) were measured by means of the 2 coaxial-end type GMX Ge detectors, GMX1 for target A and B, GMX2 for target C and D. The total muons stopped at the target calculated from the muonic x-ray spectrum was about 7$\times$10$^{9}$. 

Figure 5 shows the $\gamma$ ray spectrum measured by GMX1 after 22.25 hrs from the $\mu$ irradiation. 
In the spectrum, many $\gamma$ rays coming from RIs produced from \ce{^{Nat}}Mo($\mu$,xn) reactions can be seen and they are listed in Table 6.  
The number of RIs, N(X') produced by the MuCIP are derived from the observed $\gamma$ ray yields corrected for their decays during the $\mu$ irradiation. 
The statistical and systematical errors are included based on the Ge peak efficiency and the $\gamma$-ray branching ratio. 
The Nb isotopes with $Z$-1=41 in the large mass range of $A$=100-90 were well produced by using the natural Mo isotopes with $Z$=42 in the mass region of $A$=100-92.
Some Zr and Y isotopes with $Z$=40 and $Z$=39 were observed in mass range of $A$=97-87.
Note that similar $\gamma$ ray spectra were measured in the previous work \cite{eji13}, where isotope detection was studied.

The 140.5 keV line is the strongest $\gamma$ ray from the 6 hr isomer of \ce{^{99m}}Tc, which is the decay product of the 66 hr $^{99}$Nb produced by the $^{100}$Mo ($\mu$,1n) reaction. 
This is one of the RIs widely used as medical tracers. 
The 658.1 keV $\gamma$ ray decays from $^{97}$Nb produced by the $^{98}$Mo ($\mu$,1n) reaction shows up soon after the irradiation, but decays down quickly within 1.2 hr. 
Several $\gamma$ rays from the 23.4 hr $^{96}$Nb produced by the ($\mu$,1n) and ($\mu$,2n) reactions on $^{97}$Mo and $^{98}$Mo were observed at 569.8 kV, 778.2 keV and 1091.3 keV. 
The 765.8 keV $\gamma$ ray from the 35.0 d $^{95}$Nb is weak because of the long half-life. 
The $\gamma$-rays from the very short-lived $^{100}$Nb and $^{98}$Nb and those from very long-lived $^{94}$Nb with $1.75\times10^{8}$ hr, $^{93}$Nb $1.41\times10^{5}$ hr and $^{91}$Nb $6\times10^{6}$ hr were not observed in the present delayed spectrum. 
The yield $N(X')$ for the short-lived isotope of $^{100}$Nb is based on the previous experiment in J-PARC \cite{has18}. 
The RIs discussed so far are mainly produced by neutron emission after the $\mu$ capture.

The small peak at 743.5 keV is the $\gamma$ ray from the 16.9 hr $^{97}$Zr, which is produced mainly by the ($\mu$,1p) reaction on $^{98}$ Mo. 
The yield is about 4$\%$ of the yield of the $^{97}$Nb produced mainly by the ($\mu$,1n) reaction on $^{98}$ Mo. 
The $^{89}$Zr may be a beta decay from $^{89}$Nb as well as ($\mu$,2n1p) from $^{92}$Mo 909 keV gamma line from $^{89}$Zr is due to the beta decay.

\begin{figure}[tbp]
\begin{center}
\includegraphics[width=0.8\textwidth]{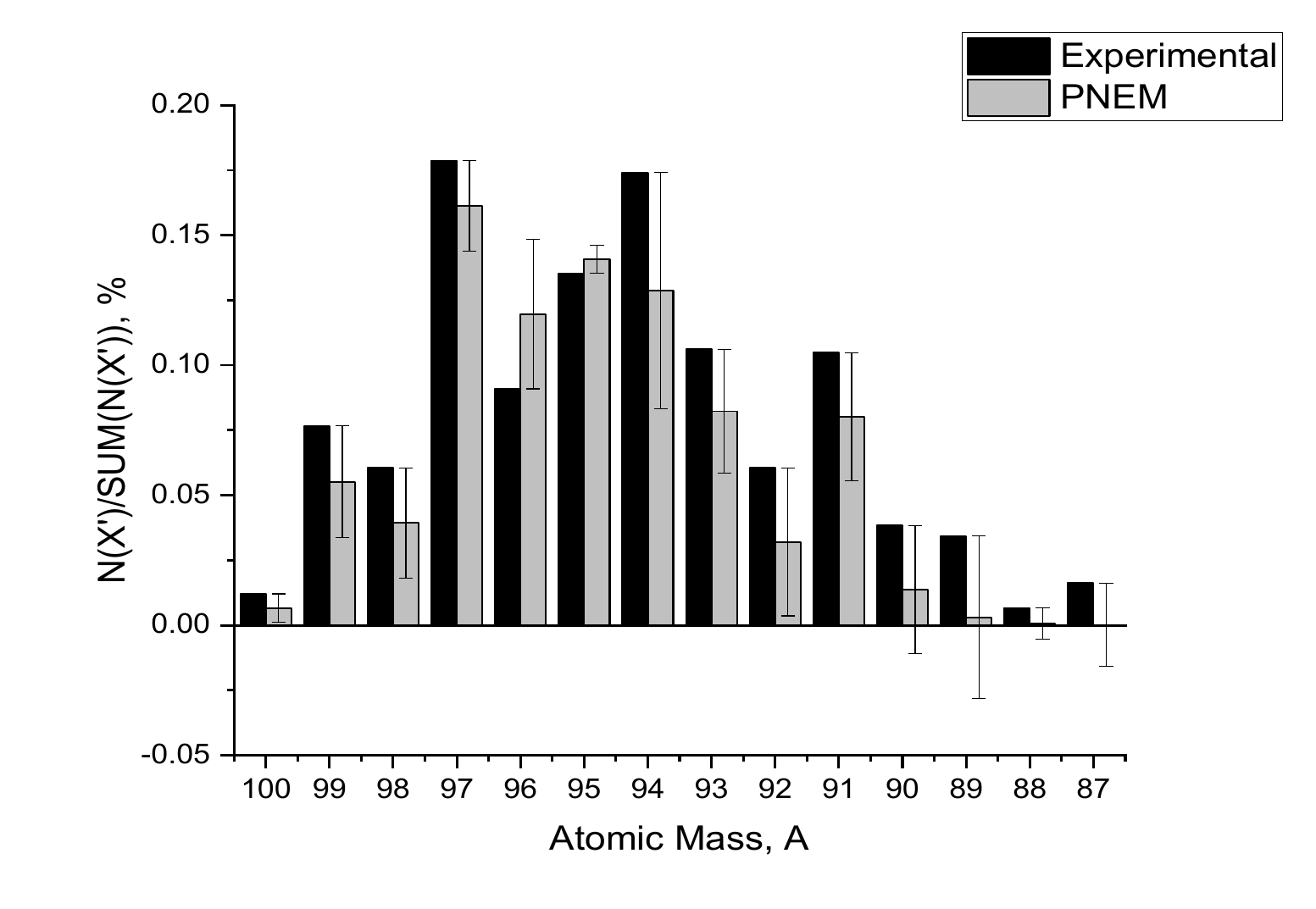}
\end{center}
\caption{Isotope mass distributions of RIs produced by MuCIP on \ce{^{Nat}}Mo.  
\label{fig:5}}
\end{figure}

The sum of the number of the observed RIs given in the Table 5 and that in Figure 6 are around 3.1$\times$10$^9$, while the number of the muons stopped at the Mo plates A and B was 3.5$\times$10$^9$.  
Thus, the total number of the RIs, including the $^{94}$Nb with $T_{1/2}$ = 2$\times$10$^4$ y and the stable $^{93}$Nb, are nearly the same as the number of the muons. 
In other words, all muons were mainly captured into the Mo nuclei to produce Nb and other isotopes.  
Here the number of protons used to produce the muons was $N_p$ = 1.45 $\times$ 10$^{16}$. 
Thus, the RI production ratio is $N_{RI}/N_{p} \approx $2.5$\times$10$^{-7}$. 
Actually, only a few $\%$ of the negative muons from the MuSIC beam line was used.  
Then, the RI production rate by using the full negative muons is around 
$N_{RI}/N_{p} \approx 7 \times 10^{-6}$.
This efficiency is very high.  
Thus one gets $N_{RI}\approx 4 \times 10^7$ per sec for 1 $\mu$ A proton beam.

Figure 6 illustrates the RI mass distribution for natural Mo based on Table 5. 
Here we clearly note that the PEQ-EQ PNEM not only reproduces the enriched target but is also able to estimate the RI fraction from experimental work.
For \ce{^{Nat}}Mo the proton emission events seem very small compared to the neutron emission events.

Based on current observations shown in Figures 3 and 5 as well as Tables 1 to 6, MuCIP can be used to produce RIs by emissions of neutrons and protons after the muon capture process. 
MuCIP RIs include several RIs for such medical (PET, SPECT) and other usages. 
The muon capture reactions to be used are 
$^{56}$Fe ($\mu$,0n)$^{56}$Mn,
$^{72}$Ge ($\mu$,0n,2n,4n) $^{72,70,68}$Ga, 
$^{90}$Zr ($\mu$,0n)$^{90}$Y, 
$^{109}$Pd ($\mu$,0n)$^{109}$Ag, 
$^{128}$Te ($\mu$,1n)$^{127}$Sb, 
\ce{^{Nat}}Pb ($\mu$,1n)$^{207,206,204,202} $Tl and others. 

\section{Remarks}
 
The present work shows that MuCIP using OMC is a very efficient and useful tool to produce various kinds of RIs. 
The isotopes produced from the target isotope of $^A_Z$X are $^{A-x}_{Z-1}$X with $x$ being 0-5. 
Among them $^{A-1}_{Z-1}$X shows a large fraction of around 50 $\%$. 
The light RIs with $Z-2$ are also produced by a proton emissions with an appreciable fraction of a few $\%$. 
MuCIP provides RIs with various lifetimes and $\gamma$-ray energies with a wide range of the mass numbers.  
They are complementary to RI productions by photon and neutron capture reactions. 
The production rate per one $\mu$ is as high as $0.5-0.1$, which is 2 orders of magnitude larger than those for photon capture reactions \cite{eji11,eji12}. 
The position (depth) of the RIs can be defined  by adjusting the $\mu$ momentum.

Muons are obtained from decaying pions which are produced by medium energy protons. 
In the case of the 400 MeV 1 $\mu$A protons from the RCNP ring cyclotron, one gets approximately 4$\times$10$^7$ muons per second at the MuSIC beam line. 
Then the RI production rate is of an order of $10^7$ per second. 
This compares well with the rate around 6$\times$10$^5$ in case of the photon capture reaction of ($\gamma $,n) by using intense photons of around 10$^8$ per sec from the NewSUBARU \cite{eji11}.  

The RI production rate by the OMC is simply proportional to the $\mu$ intensity and thus to the proton intensity.
Then one may expect high RI production rate of an order of 4 $\times$ 10$^5$ per second in case of the upgraded 10 $\mu$A proton beam at RCNP.
The PSI 0.6 GeV protons and J-PARC 3 GeV protons provide higher intensity muons of the order of 10$^{6}$ to 10$^{7}$ muons per second.  
Assuming high efficiency $\mu$ beam line as the MuSIC, then one may expect RI production rates of the same order of magnitude around the order of $10^{5}$ for 1 mA proton.
The intensity of the RIs amount to the order of 100 GBq level if one uses muons of the order of 10$^{10}$ per second. 
MuCIP is very promising future for applications in pure and applied science as part of muon capture isotope detection(MuCID) \cite{eji12}. 
Since one may use rather thin targets in the order of a few 10 mg/cm$^2$ to stop low-momentum muons, the RI density with 10$^9$ muons per sec per 10 cm$^2$ will be around 10 GBq/gr.  

We note that MuCIP is a kind of isotope transmutation process from stable to unstable isotopes, and thus is used to transform long-lived to short-lived isotopes and vice versa as well. 
The rate is of the order of 100 $\mu$ gr per year by using intense 1 mA protons from the SNS or J-PARC accelerators.  
MuCIP can be complementary to other reactions such as photo nuclear reactions \cite{eji11, szp13}, and others \cite{nag09}. 

\section*{References}

\bibliography{main}

\end{document}